# Tailoring magnetic energies to form dipole skyrmions and skyrmion lattices

S. A. Montoya[1,2], S. Couture[1,2], J. J. Chess[4], J. C. T. Lee[4,5], N. Kent[7], D. Henze[7], S. K. Sinha[3], M.-Y. Im[5,6], S. D. Kevan[4,5], P. Fischer[5,7], B. J. McMorran[4], V. Lomakin[1,2], S. Roy[5], and E. E. Fullerton[1,2] [*]

[1]*Center for Memory and Recording Research, University of California, San Diego, La Jolla, CA 92093, USA*
[2]*Department of Electrical and Computer Engineering, University of California, San Diego, La Jolla, CA 92093, USA*
[3]*Department of Physics, University of California, San Diego, La Jolla, CA 92093, USA*
[4]*Department of Physics, University of Oregon, Eugene OR 97401, USA*
[5]*Center for X-ray Optics, Lawrence Berkeley National Laboratory, Berkeley CA 94720, USA*
[6]*Department of Emerging Materials Science, DGIST, Daegu, Korea*
[7]*Physics Department, University of California, Santa Cruz, CA 94056, USA*

(Dated: December 8, 2016)

**Abstract**

The interesting physics and potential memory technologies resulting from topologically protected spin textures such as skyrmions, has prompted efforts to discover new material systems that can host these kind of magnetic structures. Here we use the highly tunable magnetic properties of amorphous Fe/Gd multilayer films to explore the magnetic properties that lead to dipole-stabilized skyrmions and skyrmion lattices that form from the competition of dipolar field and exchange energy. Using both real space imaging and reciprocal space scattering techniques we determined the range of material properties and magnetic fields where skyrmions form. Micromagnetic modeling closely matches our observation of small skyrmion features (~50 to 70nm) and suggests these class of skyrmions have a rich domain structure that is Bloch like in the center of the film and more Néel like towards each surface. Our results provide a pathway to engineer the formation and controllability of dipole skyrmion phases in a thin film geometry at different temperatures and magnetic fields.

I.    Introduction

Skyrmions are topologically non-trivial cylindrical-like magnetic domains that exhibit novel physics and potential applications to non-volatile memory [1-6]. Today, these textures exist in an array of materials from bulk magnets [7-10] to thin films [11-19] and have been shown to be stable under several physical mechanisms [19-23]. The most heavily studied mechanism to stabilize skyrmions is the Dzyaloshinskii-Moriya (DM) interaction arising in non-centrosymmetric magnetic materials or thin films with asymmetric heavy metal interface [7-14]. However, topologically similar spin structures can be stabilized by the competition of long-range dipolar energy in a thin film geometry and domain wall energy [16-19], a mechanism by which magnetic stripes and bubbles form [24-40]. Commonly a chiral magnetic bubble is termed a dipole stabilized skyrmion given the resemblance to a Bloch-type DM interaction skyrmion [2,17-19]. Given these chiral bubbles form under the application of a perpendicular magnetic field they are said to be extrinsically stable [5]. Both these class of topologically protected magnetic features are interesting and there are numerous examples of materials showing them, however, there is a limited understanding of the basic magnetic energetics required to favor their formation. These chiral bubbles or dipole-stabilized skyrmions present a test-bed to explore how the balance of ferromagnetic exchange, anisotropy and dipolar energy results in cylindrical-like domains that are topologically non-trivial.

---

[*] Corresponding author: efullerton@ucsd.edu

In this work we explore the formation of dipole-stabilized skyrmions and skyrmion lattices in amorphous Fe/Gd multilayers, with the focus on developing predictive properties that can result in the stabilization of chiral textures. Through thickness, alloy composition and temperature dependent studies of various Fe/Gd films, we find we can control the skyrmion lattice, in temperature and applied magnetic fields, by tuning the material properties of the multilayer structure. This tunability allows us to investigate the skyrmion sensitivity to material properties to manifest ordered and disordered skyrmions. By comparing experimental findings with micromagnetic modeling we show the skyrmion-lattice phase appears for a parameter space with a combination of relatively low perpendicular magnetic anisotropy ($\sim$2-4x$10^5$ ergs/cm$^3$), low magnetic moment ($\sim$350-500 emu/cm$^3$), low exchange interaction ($\sim$5x$10^{-7}$ ergs/cm), and thick films (>40 nm). Our results provide a guideline of magnetic properties required to stabilize these spin textures in thin-film ferromagnets and ferrimagnets.

The Fe/Gd films we have studied consist of multilayer deposited structures of Fe and Gd thin films (each layer <0.4nm) which are antiferromagnetic coupled forming a ferrimagnet. By appropriately choosing the thickness of the layers and the deposition conditions the films develop PMA [39-40]. In general, favorable conditions for the observation of perpendicular magnetic domains requires the uniaxial anisotropy $K_U$ be greater than the shape anisotropy $2\pi M_S^2$ where the ratio of these parameters is defined as a material's *Q*-factor [25]. For our films, the *Q*-ratio are less than 1 but by increasing the number of bilayer repetitions (*i.e.* the total film thickness) results in a transition from in-plane magnetization to the formation of perpendicular magnetic domains [41-45]. We will first show Lorentz transmission electron microscopy (TEM), resonant soft x-ray scattering and full-field transmission X-ray microscopy to confirm the presence of skyrmions and skyrmion lattices in our films and determine the sensitivity of the skyrmion formation to the temperature, applied magnetic field range and film thickness. We will then quantify the materials parameters of our films and compare the experimental results to micromagnetic simulations.

II.     **Methods.**

The Fe/Gd specimens are sputter deposited at room temperature in a UHV environment under a 3mTorr Argon environment. To grow the film structures, we alternatively deposit Fe and Gd layers of a specific thickness, and continue process until achieving multilayer with the desired number of layers. Films have a seed/capping layer of Ta 5nm to protect the films from corrosion. Samples are deposited on a range of different substrate for magnetic and imaging characterization, including 50nm and 200nm SiN membranes and Si substrate with a native oxide layer.

The field and temperature dependence of the magnetic domain morphology is imaged using a variety of techniques, which include: (*i*) Lorentz TEM using an FEI Titan in Lorentz mode equipped with an image aberration corrector, (*ii*) soft x-ray scattering (SXS) at the Gd $M_5$ (1198eV) absorption edge and Fe $L_3$ (708eV) absorption edge at Beam-line 12.0.2 Advanced Light Source, Berkeley National Lab and (*iii*) full-field transmission X-ray microscopy along the Fe $L_3$ (708eV) absorption edge performed at Beam-line 6.1.2 Advanced Light Source, Berkeley National Lab.

Magnetic hysteresis measurements were characterized using a Quantum Design PPMS cryostat with vibrating sample magnetometer and a magneto-optical kerr effect magnetometer. The ferromagnetic resonance measurements were performed using a custom probe that affixes a coplanar waveguide at one end and can be inserted into a Quantum Design PPMS for temperature and d.c. magnetic field $H_{dc}$ probing. The *r.f.* field $h_{rf}$ lies in the plane to magnetic film specimen and $H_{dc}$ is transverse to the coplanar waveguide. The ferromagnetic resonance is extracted from the reflection coefficients of scattering parameters using an Agilent PVNA E8363B in field-swept mode.

## III. Results

### A. Skyrmion lattice formation at room temperature.

Figure 1 shows the field dependence of the magnetic domain morphology of a [Fe (0.34 nm)/Gd (0.4 nm)]x80 multilayer (total thickness 53.6nm, see Supplementary Section C) imaged by means of Lorentz TEM at room temperature. The under-focused images are obtained as a magnetic field perpendicular to the sample is swept from zero towards magnetic saturation. Each field image is from the same region of the sample but not the same specific area of the sample as the image position shift with applied magnetic field. At zero field the film exhibits stripe domains that have a periodicity of ~124 nm (Fig. 1a) and random in-plane order. The images reveal the in-plane magnetization orientation averaged over the thickness of the magnetic domains through variations of darkness/brightness intensity, and domain walls are evidenced by strong and sharp dark/bright contrast changes. Analysis of over/under-focused Lorentz TEM images using the transport-of-intensity equation [46, 47] allows us to determine the projected in-plane magnetic induction, shown as a color map in Fig. 1b. Here, the color indicates the direction of the magnetic induction with respect to the color-wheel insert; in a similar fashion, the color intensity details the magnitude of the magnetic induction. The images show that the stripe domains are separated by Bloch walls (Fig. 1b) where the in-plane moment of the walls is parallel to the stripes. This arrangement is more clearly observable in the enlarged image (Fig. 1c) detailing the enclosed region of Figs. 1a and 1b, where both the color and vector maps detail the direction of the Bloch wall. Since the films are relatively thick and no DMI is anticipated a Bloch-like domain structure is expected. As will be discussed below there are likely closure domains (*e.g.* Néel caps) at the top and bottom of the films [48]. Evidence of closure domains will not be visible in these images since the orientation of these domains at the top and bottom of the films are opposite in direction and will average to zero in Lorentz TEM images.

As a magnetic field is applied, perpendicular to the film (Figs. 1d and 1e), the stripes with magnetization parallel to the field grow at the expense of domains opposite to the field. We observe that the stripe domains whose magnetization is opposite to the field begin to collapse into cylindrical domains (Figs. 1d and e). Since the Bloch-line continuously wraps around the cylindrical domain it is defined as a skyrmion with winding number S = 1 [2,3]. If the Bloch-line wraps continuously in a clockwise fashion then it has helicity $\gamma = -\pi/2$, whereas if the Bloch-line wraps counter-clockwise direction it has helicity $\gamma = +\pi/2$ [2]. At $H_z$ = 1450 Oe, the skyrmions have a diameter of ~71nm and their size does not vary significantly from initial formation to annihilation. The enclosed/enlarged region, detailed by red boxes in Figs. 1d and 1e, shows the extremities of a stripe domain collapsing into a skyrmion (Fig. 1f), as well as, as the color/vector map representation of the magnetic textures enclosed. At this magnetic field strength, the domain morphology consists of a combination of disordered stripe domains and skyrmions (Figs. 1c and 1d). As the magnetic field is further increased, the entire film fills with dipole skyrmions with an equal population of the two possible helicities (Figs. 1g and 1h). The two different helicities ($S=1$, $\gamma = \pm\pi/2$) appear as dark and light cylindrical textures in Fig. 1g and they arrange into a weakly-ordered hexagonal lattice (Figs. 1g and 1h). This skyrmion lattice is stable for a wide range of magnetic fields spanning from 1700 to 2400 Oe. By increasing the magnetic field further, the skyrmion lattice dissociates into a disordered isolated skyrmion state where these textures begin to collapse as the film reaches magnetic saturation (Figs. 1j and 1k).

**B. Temperature dependence of the skyrmion phase.**

Using a combination of real and reciprocal space imaging techniques we explored the dependence on temperature and applied magnetic fields that result in the formation of the skyrmion phase for two Fe-Gd compositions. From scattering patterns obtained by means of resonant soft x-ray scattering, see Supplementary Section A, we identified four regions emphasizing long-range ordered magnetic states including: (*i*) disordered stripe domains, (*ii*) a stripe-to-skyrmion transition, (*iii*) a skyrmion lattice and (*iv*) uniform magnetization. Then using images obtained from Lorentz TEM and transmission soft x-ray microscopy we supplemented our findings in the magnetic phase map. From real-space images, we identified an additional region detailing (*v*) disordered skyrmions that occurs after the skyrmion lattice dissociates. Since disordered skyrmions do not have any long-range order, this region is not easily determined in scattering experiments.

Figure 2 summarizes the various magnetic domain configurations observed when an applied field, perpendicular to the film applied from zero-field to magnetic saturation at detailed temperatures from 300K to 50K. We find that the [Fe (0.34 nm) /Gd (0.4nm)]x80 multilayer exhibits a similar domain morphology, as previously described with Lorentz TEM, with a skyrmion lattice that extends from room temperature to 220K for a wide range of applied magnetic fields (Fig. 2a). At temperatures where no skyrmion phase exists, the Fe/Gd film primarily exhibits disordered stripe domains. In the case we modify the Fe-content of the Fe/Gd structure to [Fe (0.36nm) /Gd (0.4nm)]x80, we find a similar broad skyrmion phase that is shifted to lower temperatures, spanning a temperature window from 180K to 100K (Fig. 2b). Here, it is evident that the formation of a skyrmion phase is sensitive to the Fe/Gd composition and is tunable to a temperature range of interest.

### C. Thickness dependence on stabilizing skyrmions.

To further explore the formation of a skyrmion lattice at room temperature, we varied the number of Fe/Gd bilayer repetitions in the multilayer to investigate the role of the thin film magnetostatic energy on the skyrmion phase. The multilayers studied consisted of [Fe (0.34nm)/ Gd(0.41nm)]$_{xN}$ where N = 40, 80 and 120 repetitions where only the latter two film structures exhibited evidence of perpendicular magnetized stripe-like domains in their magnetic hysteresis loops (see Supplementary Section B); whereas [Fe (0.34nm)/ Gd(0.41nm)]$_{x40}$ appears consistent with in-plane magnetic domains. This suggests that as the number of bilayers is increased, we gain magnetostatic energy to form perpendicular magnetic domains. Similar observations of a thickness driven spin reorientation of the magnetization, from in-plane to perpendicular, with increasing thickness have has been reported for numerous systems including hcp(0001) Co films [42, 43].

When examining the field-dependence of the domain morphology by means of transmission soft x-ray microscopy measured at the Fe L$_3$ (708eV) absorption edge we verified that N=40 film does not show any perpendicular magnetic domains while the N=80 and N=120 repeat films exhibited both stripe and skyrmion magnetic domain textures consistent with Fig. 1. The [Fe (0.34nm)/ Gd(0.41nm)]$_{x80}$ film exhibits stripe domains at zero applied field where the contrast is sensitive to the out-of-plane magnetization (unlike Lorentz TEM which is sensitive to the in-plane magnetization) (Fig. 3a). Here domains with magnetization parallel/anti-parallel to the perpendicular direction appear as dark/white textures. The presence of a small remanent in-plane magnetic field in the measurement causes the stripe domains to align in the direction of this field. With a modest out-of-plane field ($H_z$ = 500, 625 Oe), we observe the stripes begin to pinch into cylindrical textures that occupy the same physical space of the original stripe (Fig. 3b-c). At this field there is a near equal population of stripes and aligned cylindrical domains. When the field is further increased ($H_z$ = 750 Oe), we find all the stripes have pinched into aligned cylindrical textures that do not arrange into a close packing lattice (Fig. 3d). As the field is further increased, the cylindrical domains dissipate (Fig. 3d) and we are left with the cylindrical textures that first formed (Fig. 3b-c). This suggests that we observe two different magnetic domains as a consequence of the in-plane field. In one case, the Bloch-line of most stripes aligns in the direction of the in-plane field which results in the formation of magnetic bubbles (zero-chirality) when a perpendicular field is applied as described in Ref. 16; whereas, stripes with random Bloch-line arrangement result in skyrmions (chiral textures) as we have previously shown. Specifically, we have shown a skyrmion molecule [16] is made up of a bound pair of opposite helicity skyrmions which can also be stabilized in Fe/Gd films as a result of applying a fixed in-plane field and then applying a perpendicular magnetic field.

The thickest film studied, [Fe (0.34nm)/Gd(0.41nm)]$_{x120}$, showed disordered stripe domains (white domains) and dumbbell domains (dark domains) at zero-field. Here the remanent in-plane field is insufficient to cause the stripes to order, as observed previously. When a perpendicular field is applied, the dumbbell domains begin to merge to form disordered stripes (Fig. 3h) and by $H_z$ =750 Oe the stripes with magnetization opposite to the field begin to collapse into skyrmions (Fig. 3i). By slightly increasing the field again, all the stripes have collapsed into skyrmions and these have arranged into a weakly coupled hexagonal lattice which spans from $H_z$= 800 Oe to 2500 Oe (Fig. 3j-l). We find the skyrmions have a diameter of ~70nm and that the size does not vary significantly from their initial formation to annihilation. The variation of film thickness, clearly demonstrates that skyrmions, like bubbles, require a specific ratio of magnetic properties ad film thickness for these magnetic textures to become favorable [25,27].

**D. Magnetic and ferromagnetic resonance properties.**

To determine the magnetic properties of the films in which skyrmion lattice formation become favorable (Figs. 2a and b) we performed temperature dependent magnetometry and ferromagnetic resonance (FMR) measurements. From magnetic hysteresis loops (Supplementary Section B) measured both by magnetometry and magneto optical Kerr effect measurements, we determine that the Fe/Gd ferrimagnetic moment is Gd rich for these films at all measurement temperatures. This is consistent with measurements of bulk alloys with similar compositions. Both Fe/Gd films show relatively low magnetic moment that varies similarly with decreasing temperature (Fig. 4a). The slightly stronger dependence of M$_S$ *vs*. T for the [Fe(0.34nm)/Gd(0.4nm)]x80 is also consistent with the lower Fe content that lead to a higher low-temperature moment and lower $T_C$ compared to the higher Fe content films. At temperatures where the Fe/Gd films exhibit a skyrmion phase, the magnetic moments are similar (Fig. 4a) thus does not explain the shift of the skyrmion phase to a different temperature range (Fig. 2).

To characterize the anisotropy FMR measurements were performed with perpendicular fields above magnetic saturation (Supplementary Section B). The resonance fields vary linearly with applied magnetic field as expected from the Kittel formula [49], $f = \frac{\gamma_e}{2\pi}(H_{dc} - 4\pi M_S + H_k)$ where $\gamma_e$ is the $e^-$ gyromagnetic ratio, $H_{dc}$ is the applied field, $4\pi M_S$ is the demagnetization field, $ok$ is the saturation magnetization and $H_K = 2 \cdot K_U/M_S$ is the perpendicular anisotropy field where $K_U$ is the uniaxial perpendicular anisotropy that is developed in the thin-film growth process. The intercept with the field axis when frequency becomes zero is given when $H_{dc} = 4\pi M_S - H_k = 4\pi M_{eff}$. For all of our films the intercept is positive (Supplementary Section B) indicating a relatively weak induced perpendicular anisotropy such that $K_U < K_D = 2\pi M_s^2$ or $Q = K_U/K_d < 1$ so the effective anisotropy is in-plane. The extracted values for $M_S$, $4\pi M_{eff}$ and $K_U$ are given in Figs. 4a-c, respectively, as a function of temperature. As the temperature is reduced, both $M_S$ and $4\pi M_{eff}$ increase roughly linearly with temperature which suggests the films become more in-plane (Fig. 4a, b) while a modest decrease of the intrinsic perpendicular anisotropy is observed with decreasing temperature (Fig. 4d). This atypical temperature dependence of the intrinsic anisotropy has been previously reported for RE-rich Fe/Gd films with a small bilayer periodicity $t_{Fe} + t_{Gd} \leq$ 1.4nm (the Fe/Gd films detailed in this work have a period < 0.8nm) [50,51]. The increase in anisotropy with increasing Fe layer thickness is observed in a series of films and appears to be a general feature of the Fe/Gd system. We have further indicated in Fig. 4 the regions in temperature where we observe the skyrmion lattice phase.

The fact that domains and skyrmion lattices are only observed for film thickness above a critical thickness is consistent with earlier studies of films where $K_U < K_d$ [41-45]. There are theoretical estimates for the critical thickness for the onset of weak stripe domains given by [Refs. 42, 44]:

$$t_1 \sim 17.7\sqrt{A}\, M_S^2/K_U^{3/2} \qquad (1)$$

where $A$ is the exchange parameter. The predicted value for the thickness for the magnetization rotate from in plane to out of plane with stripe domains is given by [Refs. 29, 42, 45]:

$$t_2 \sim 27.2\sqrt{A}\, M_S^2/K_U^{3/2}. \qquad (2)$$

In both cases the critical thickness scales as $M_S^2/K_U^{3/2}$. In Fig. 4d we plot this ratio as a function of temperature. We see this ratio increases with decreasing temperature reflecting the strong temperature dependence of the saturation magnetization. We also find that we experimentally observe the skyrmion phase for the same range of this ratio from $0.735 \times 10^{-3}$ to $1.359 \times 10^{-3}$ as shown in Fig. 4d. While neither the values of $M_S$ or $K_U$ appear predictive for determining the temperature range where we observe the skyrmion the ratio as $M_S^2/K_U^{3/2}$ does provide a guideline to the formation of the skyrmion phase for the Fe/Gd system.

The only material parameter in Eqs. (1) and (2) we do not have a quantitative measure of is the exchange stiffness constant $A$, however, a qualitative measure can be deduced from noting that no super-lattice peaks are observable in small angle x-ray reflectometry (Supplementary Section C). This suggest that there is strong intermixing between the Fe and Gd layers; as a consequence, we can assume the Fe/Gd films are layered alloy-like structures that resemble an Fe-Gd alloys of similar composition. For this reason, we assume an exchange stiffness $A$ between 2 - $5 \times 10^{-7}$ erg/cm based on values reported for similar Fe-Gd alloys [52-55]. Using these values of $A$ we would predict a $t_1$ in the range 50.2 nm $< t_1 <$ 79.4 nm in reasonable agreement for the transition from in-plane to stripe phase between 40 and 80 repeats as observe experimentally.

### E. Micromagnetic modeling

To understand the mechanism stabilizing the skyrmion phase, we performed numerical simulations of the Landau-Lifshitz-Gilbert (LLG) equation, utilizing the FASTMag [56] solver for a magnetic slab that is 2000 nm x 2000 nm x 80 nm. The input magnetic parameters were obtained from experiment (for further details see Figure captions). Shown in Fig. 5 is the field evolution of equilibrium states that results for $M_S = 400$ emu/cm$^3$, $K_U = 4 \times 10^5$ erg/cm$^3$ and $A = 5 \times 10^{-7}$ erg/cm. These values are within the range of magnetic properties of both [Fe (0.34nm) /Gd (0.4nm)]x80 and [Fe (0.36nm) /Gd (0.4nm)]x80 (Fig. 4). Initially the slab is saturated along the z-axis and then the perpendicular field is reduced to zero-field. At any field step the slab is allowed 30 ns to relax into an equilibrium state. Figures 5a (top view) and 5b (side view) are the results at zero-field that exhibit a configuration of disordered stripe domains similar to those observed in Fig. 1. The typical domain periodicity is ~183 nm. The cross-section view of the magnetization across the film thickness along the line given in Fig. 5a reveals a Bloch-like wall configuration at the center of the slab (Fig. 5b $m_y$ component) while near the top and bottom of the film there are flux closure caps (Fig. 5b, $m_x$ component). This domain arrangement is expected given the low $Q$-factor ($Q \sim 0.4$) as previously determined by resonant x-ray scattering from Fe/Gd films [48].

When a magnetic field is applied perpendicular to the slab, the stripe domain with magnetization opposite to the field direction collapse into individual skyrmions (Fig. 5c-e). As the magnetic field is increased further, the chiral textures arrange into a hexagonal lattice that exists from $H_{dc}$ = 1700 Oe to 2300 Oe (Figs. 5f, g, n-r). The typical skyrmion size is ~83 nm and the separation is given by ~128 nm at $H_{dc}$ = 1700 Oe. As the magnetic field is increased, the skyrmion size decreases to 53 nm at $H_{dc}$ = 2700 Oe. The skyrmion size, separation and field history are in good agreement with our experimental observation. For Figs. 5a, c-g and n-r we plot the $m_z$ component for magnetization which can be compared to the contrast in Fig. 3.

For the skyrmions that form in Fig. 5 the domain walls are Bloch-like in the center of the film and we find a roughly 50% percent chance of the two helicities. This can be seen in Fig. 5g where we plot the $m_x$ component of magnetization at the center of the slab for $H_{dc}$=1700 Oe. The skyrmions where $m_x$ component is red above the skyrmion and blue below have one circulation of the domain wall and we characterize it by a winding number $S = 1$, $\gamma = -\pi/2$. The skyrmions where $m_x$ component is blue above the skyrmion and red below have the opposite circulation and we characterize it by a winding number $S = 1$, $\gamma = \pi/2$. These results agree with the Lorentz TEM images of Fig. 1. By tracking the orientation of the in-plane components of the domain walls in Fig. 5 with increasing applied field we find that the chirality of the skyrmions is determined by the chirality of the walls of the original domain. There are also examples of bubbles that form where the $m_x$ component is either red or blue both above and below the bubble and we characterize this by a winding number $S = 0$. Such bubbles are also observed in Lorentz TEM experiments.

Shown in Fig. 5(h) is the magnetic projection through the thickness of the film for the dashed line shown in Figs. 5(f) and 5(g). The $m_z$ projection shows that the core of the skyrmion extends through the thickness of the film but narrows slightly near the top and bottom of the film. The $m_y$ projection shows the circulating Bloch walls around each skyrmion and the $m_x$ projection shows the existence of closure domains at the top and bottom of the films (this is shown in more detail in Fig. 5i-m). These closure domains will not be seen in the experimental images in Figs. 1 and 3 because both Lorentz TEM and transmission x-ray microscopy average over of the thickness of the film. As seen in Fig. 5(h) the closure domains have the opposite sign at the top and bottom of the films and will average out in a transmission experiment.

The details of the domain-wall structure for one $S = 1$, $\gamma = -\pi/2$ skyrmion are shown in Fig. 5(i-m). Shown are slices for different depths within the slab where the color gives $m_z$ and the arrows give the direction of the in-plane magnetic component and length corresponds to the magnitude of $m_x$ and $m_y$. At the center of the slab (z = 0 nm), $m_x$ and $m_y$ continuously wrap around the skyrmion forming a Bloch wall and the wall width is the narrowest. As one progresses towards the top (z = 40 nm) or bottom (z = -40 nm) surface you see the domain wall broadens and becomes more Néel like. At the top surface the in-plane magnetization points mostly radially in forming the Néel caps. At the bottom surface the in-plane magnetization now points away from the skyrmion center. For a skyrmion with opposite helicity, the wrapping of $m_x$ and $m_y$ toward the center of the skyrmion core is inverted; while, the Néel cap configuration at the top and bottom of the slab is the same.

For higher fields the skyrmions become disordered and begin to disappear, first at the edges and then throughout the film as the field approaches magnetic saturation (Figs. 5n-r). Overall, the field dependent domain morphology is in good agreement with our experimental observations. Here the micromagnetic model suggests the chiral bubble domains form due to minimization of competing demagnetization energy and domain wall energy, and that thermal fluctuations are not required for their formation since this is a zero-temperature model. To compare our numerical results to our experimental observations of chiral cylindrical textures we computed the magnetization projection averaged over the film thickness. This is shown in Fig. 6 for $<m_z>$ and $<m_x>$ and are compared to selected Lorentz TEM and x-ray images. There is no signature of the different helicities of the $<m_z>$ as expected and consistent the x-ray images. The $<m_x>$ and $<m_y>$ projection is only sensitive to the Bloch nature of the wall as seen in Lorentz TEM.

Next, we discuss the magnetic properties required to stabilize a dipolar field driven skyrmion phase. As we have demonstrated the composition of the Fe/Gd specimen directly correlates the temperature and applied magnetic field range in which skyrmions become favorable particularly a window of $M_S^2/K_U^{3/2}$ values. To understand the effect of exchange $A$ in the formation of a skyrmion phase we simulated magnetic domains similar to those in Fig. 5 for a fixed applied field $H_z$ = 2000 Oe and $M_S$ = 400 emu/cm$^3$ and various $K_U$ and $A$ values are shown in Fig. 7(a). We find that modest changes in either of these parameters leads to different equilibrium states. For instance, as $K_U$ is increased, the skyrmion lattice quickly becomes less correlated and the skyrmion size begins to vary. Increasing the exchange $A$, for any fixed $K_U$ except $K_U$ = 2x10$^5$ erg/cm$^3$, results in a skyrmion arrangement that becomes more disordered. For a fixed $M_S$, a close packing lattice of skyrmions is only achievable in a narrow region of both $K_U$ and $A$. Our modeling suggests a modest value of the exchange parameter is a critical parameter that determines the long-range order of skyrmions forming a lattice, and supports the existence of a weak exchange in these Gd-rich Fe/Gd films.

## IV.    Discussion

Using a combination of numerical simulations and experimental data, we have constructed a $K_U$-$M_S$ phase map (for a fixed $A$ = 5x10$^{-7}$ erg/cm) that illustrates the measured and calculated magnetic domains at zero-field and applied field values (Fig. 7b). In the phase map, the $Q$-ratio is also plotted in the foreground with contour lines depicting variations in $Q = K_U/2\pi M_S^2$ for domain states with $Q < 1$. The $Q$-ratio serves as a heuristic to estimate volume fraction distributions of perpendicular domains and Néel caps in these thick Fe/Gd films. We recall that when films exhibit a $Q$-ratio greater than 1, the domain morphology favors perpendicular domains with negligible Néel caps; likewise, as $Q$ decreases below 1 the volume fraction of the Néel caps grows and the perpendicular domains occupy a lower volume fraction.

Inspecting the domain states when the anisotropy is very low ($K_U \leq$ 2x10$^5$ erg/ cm$^3$) such that the thickness is below $t_1$ (Eq. 1) we find the magnetization is in-plane and as a result of the geometry of the slab, the system favors the formation of an in-plane vortex configuration. For larger $K_U$ (or lower $M_S$) the film transitions to out-of-plane magnetic domains. The two predominant magnetic configurations consist of labyrinth domains at zero-field that form skyrmions when a field is applied perpendicular to the slab. What differentiates the regions are the mechanisms by which the skyrmions form, in one case: (i) the stripes pinch-off into chiral bubbles as seen in Fig. 5, and in the other (ii) the extremities of the stripe collapse to form single chiral bubbles that do not arrange in a lattice as seen in Co/Pt multilayers [57] for instance. For only a small region of $K_U$ and $M_S$ values do we observe the skyrmions arranging in a close packing lattice; in addition, this region exists at $Q$-value ratios from 0.2 to 0.4 which suggests the overall domain structure of perpendicular domains and Néel caps is fairly comparable. As we move away from the skyrmion lattice region, the distance between the cylindrical textures increases until they become disordered.

In the foreground of Fig. 7b we also detail the range of $M_S^2/K_U^{3/2}$ values where an ordered skyrmion phase is observable numerically and experimentally. Equilibrium states within this region of varying $M_S$ and $K_U$ will share similar critical thickness $t_1$, for a fixed $A$, at which weak perpendicular stripe domains will form, but only a small region of low $M_S$ and $K_U$ results in a close packing lattice of skyrmions. Large $M_S$ and $K_U$ result in stripe domains that form disordered chiral bubbles. This suggests that the critical thickness is not a sole determinate of ordered skyrmions.

The Lorentz-TEM images (Fig. 1) and numerical simulations (Fig. 5, 6) suggests the stabilization of these skyrmions is purely driven by competing dipolar and exchange energies and that no DMI is present in these films. The Lorentz-TEM images show two helicity textures with an equal population distribution in the skyrmion phase. If some DMI were present, then the system would likely favor the formation of a chiral domain compared to the other, as well as a Néel cap orientation over the other which is not the case here. The fact that we numerically observe the stabilization of the same 2-helicity skyrmions on a slab with no DMI supports this observation. Given the nature of these skyrmions, these films could potentially also be designed to host antiskyrmions as recently theoretically predicted [58]. Unlike bubble domains which typically observed in materials with $Q > 1$ and exhibit strong PMA, our chiral cylindrical domains appear in a material parameter space where $Q < 1$ and the formation of perpendicular domains results from a thickness driven domain morphology rearrangement [41-45].

In conclusion, we have demonstrated the existence of skyrmion lattice in Fe/Gd films by means of real and reciprocal space imaging techniques. We have shown that by tuning the magnetic properties and film thickness we can control the stabilization of a skyrmion phase in temperature and applied magnetic fields. The simplicity of the magnetic material and the easily tunable properties makes it of interest as a potential test bed for studying physics of dipole skyrmions, as well as, for potential memory technologies. Furthermore, the universality of our numerical model presents a roadmap to design new classes of materials that can exhibit dipolar field driven skyrmions.

**Acknowledgements**

Work at UCSD including materials synthesis and characterization, participation in synchrotron measurements and modeling was supported by U.S. Department of Energy (DOE), Office of Basic Energy Sciences (Award No. DE-SC0003678). Work at University of Oregon was supported by the U.S. Department of Energy (DOE), Office of Science, Basic Energy Sciences (BES) under Award # DE-SC0010466. Work at the ALS, LBNL was supported by the Director, Office of Science, Office of Basic Energy Sciences, of the U.S. Department of Energy (Contract No. DE-AC02- 05CH11231). S.A.M. acknowledges the support from the Department of Defense (DoD) through the Science, Mathematics & Research for Transformation (SMART) Program.

B.J.M. and J.J.C. gratefully acknowledge the use of CAMCOR facilities, which have been purchased with a combination of federal and state funding. M.Y. Im acknowledges support by Leading Foreign Research Institute Recruitment Program through the National Research Foundation (NRF) of Korea funded by the Ministry of Education, Science and Technology (MEST) (2012K1A4A3053565 and 2014R1A2A2A01003709). S.K. and P.F. acknowledge support by the Director, Office of Science, Office of Basic Energy Sciences, Materials Sciences and Engineering Division, of the U.S. Department of Energy under Contract No. DE-AC02-05-CH11231 within the Nonequilibrium Magnetic Materials Program (KC2204) at LBNL.

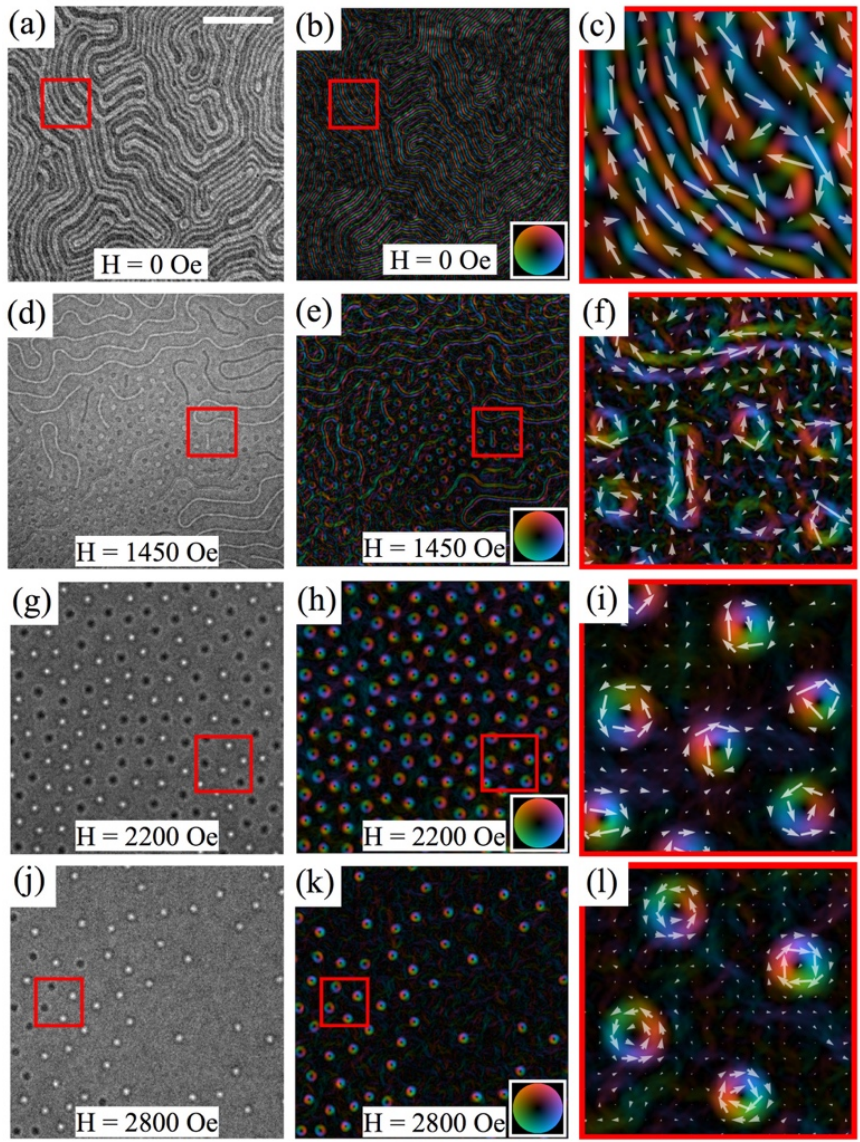

**Figure 1. Real space imaging of the field-dependent magnetic domain morphology of [Fe (0.34 nm)/Gd (0.4 nm)]x80.** Under-focused Lorentz TEM images (first column) measured at room temperature and their corresponding magnetic induction color maps (second column) are detailed. The images are captured as a perpendicular magnetic field is applied from zero-field to magnetic saturation. Four different magnetic states are observed as the field is swept, including: disordered stripe domains (**a, b**), stripe-to-skyrmion transition (**d, e**), skyrmion lattice (**g, h**) and disordered skyrmions (**j, k**). Enclosed regions in the first two columns are enlarged to detail the in-plane magnetic domain configuration using both color and vector magnetic induction maps in the third column (**c, f, i**). The scale bar in **a** corresponds to 1μm.

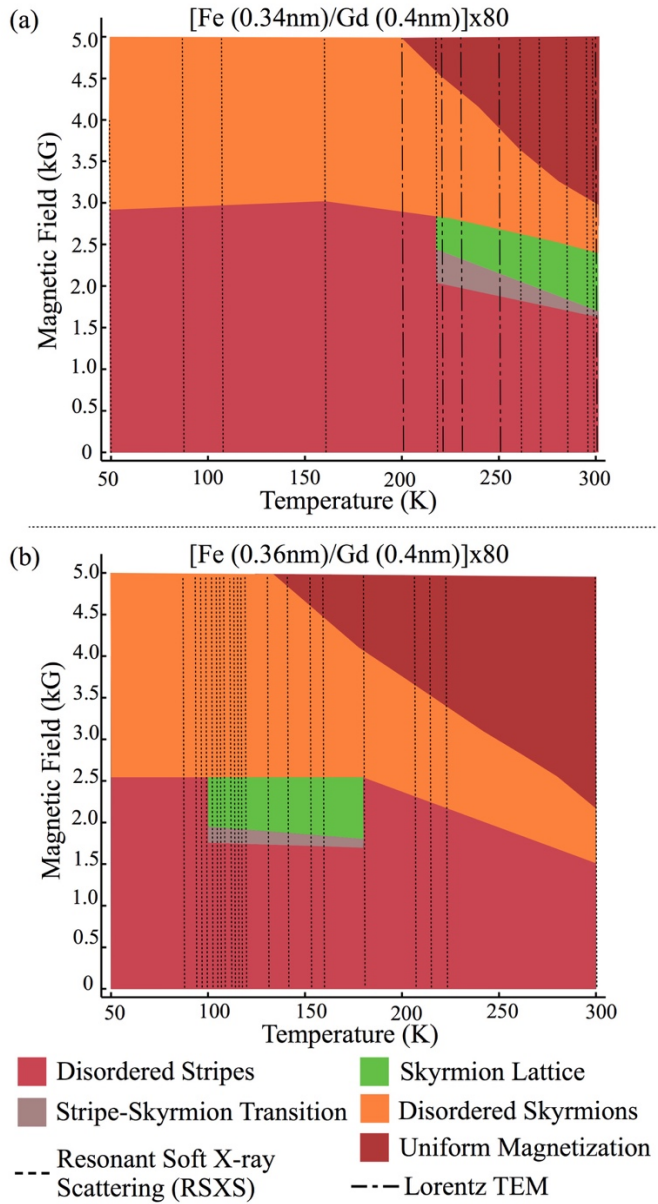

**Figure 2. Magnetic field and temperature dependence of the skyrmion phase.** The magnetic phase diagram for two Fe/Gd film structures are shown: **(a) [**Fe (0.34 nm)/Gd (0.4 nm)]x80 exhibits a broad skyrmion phase around room temperature, and **(b) [**Fe (0.36 nm)/Gd (0.4 nm)]x80 shows a similar skyrmion phase that is shifted to lower temperatures. These magnetic phase maps are constructed using data from four different imaging techniques: resonant soft X-ray scattering, Lorentz TEM and transmission X-ray microscopy (at room temperature only). The marker-lines detail the temperature and imaging technique used to scan the domain morphology.

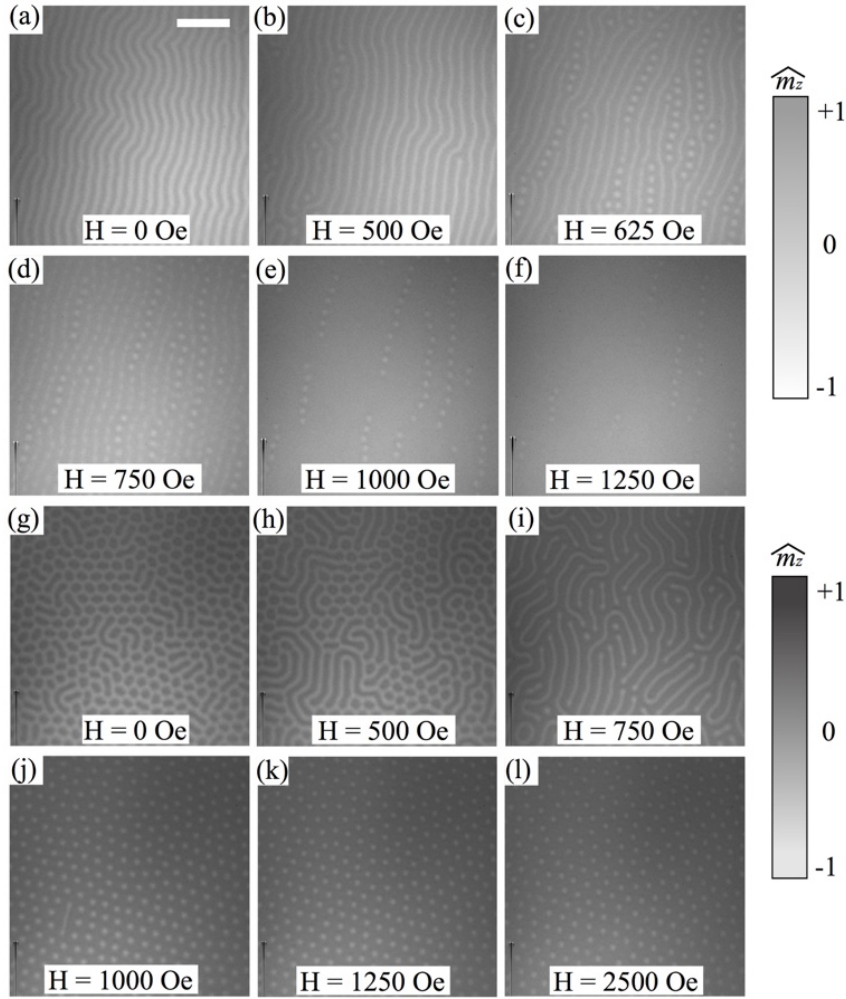

**Figure 3. Film thickness dependence of [Fe (0.34 nm)/Gd (0.41nm)]xN.** The domain morphology, obtained by X-ray microscopy, for Fe/Gd multilayers with different bilayer repetitions are detailed, as a perpendicular magnetic field is applied from zero-field to magnetic saturation. **(a-f)** [Fe (0.34nm)/ Gd(0.41nm)]$_{x80}$ exhibits stripe domains at zero-field **(a)** that pinch off into skyrmions as the magnetic field is increased **(b-c)**. Above $H_z = 750$ Oe the cylindrical textures begin to collapse in aligned clusters **(e-f)** until no skyrmions can be observed in the field of view. **(g-l)** [Fe (0.34nm)/ Gd(0.41nm)]$_{x120}$ exhibits disordered stripe domains at zero field **(g)** that collapse begins to collapse into skyrmions as the magnetic field is increased **(h)**. These skyrmions arrange into closed-pack hexagonal lattice from $H_z = 800$ Oe to 2500 Oe **(j-l)**. The scale bar in **a** corresponds to 1μm.

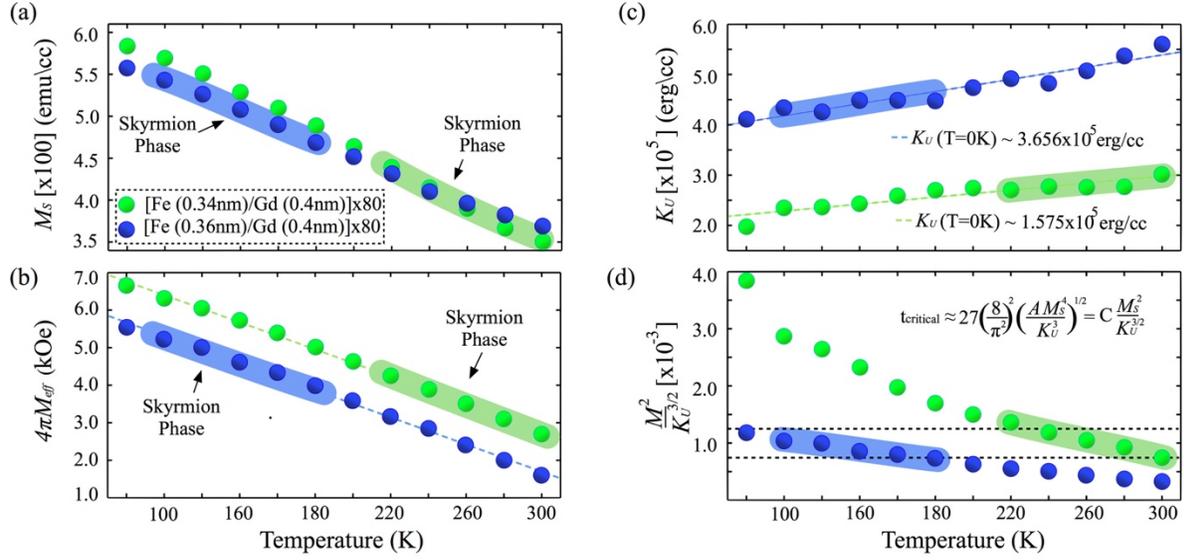

**Figure 4. Temperature dependent magnetic and ferromagnetic resonance properties.** The magnetization **(a)**, effective magnetization **(b)**, uniaxial anisotropy **(c)** and the $M_S^2/K_U^{3/2}$ ratio **(d)** are shown for **[Fe (0.34nm)/Gd (0.4nm)]x80** and [Fe (0.36nm)/Gd (0.4nm)]x80. The region in temperature where each Fe/Gd film exhibits a skyrmion phase has been shaded to serve a guide to the eye. By linearly fitting the uniaxial anisotropy with temperature, we can conclude that $K_U$ is positive at absolute-zero (**c**). The enclosed region with dashed lines in **(d)** represents a window of $M_S^2/K_U^{3/2}$ values were the both Fe/Gd films exhibit a skyrmion phase.

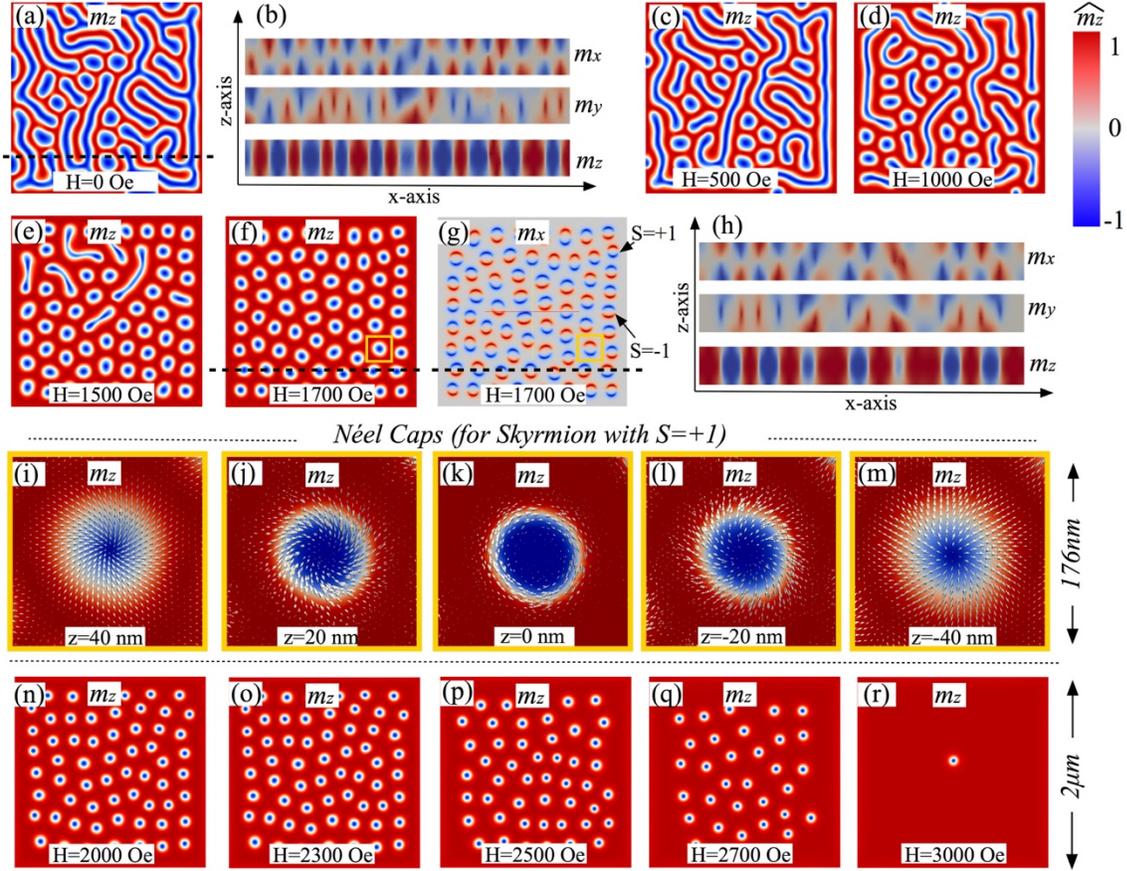

**Figure 5. Micromagnetic modeling of domain morphology. (a-r)** [$M_S$ = 400 emu/cm$^3$, $K_U$ = 4x10$^5$ erg/cm$^3$ and $A$ = 5x10$^{-7}$ erg/cm] The equilibrium states illustrate the field dependent domain morphology at several magnetic fields that capture the domain evolution from a stripe to a skyrmion phase. These equilibrium state primarily depict the top side view of the magnetization along the z-axis ($m_z$) at the top surface of the slab (z=40nm). The magnetization ($m_z$) is represented by regions in red (+$m_z$) and blue (-$m_z$); whereas the in-plane magnetization ($m_x$, $m_y$) is represented by white regions surrounding the blue features. **(b, h)** Illustrates the lateral magnetization components ($m_x$, $m_y$, $m_z$) across the film thickness for the disordered stripe domains in **(a)** and the skyrmion phase in **(f, g)** along the dashed line. Inspection along the lateral magnetization reveals a Bloch-like wall configuration with closure domains in both states. The chirality of the skyrmions is depicted in **(g)** along top side-view of $m_x$ across the center of the slab. **(i-m)** Detail the magnetization distribution at different depths (z = 40, 20, 0, -20, -40nm) for a skyrmion with chirality S = +1, $\gamma$ = -$\pi$/2 that is enclosed in a box in **(f, g)**. At each depth, the perpendicular magnetization is represented by blue (-$m_z$) and red (+$m_z$) regions and the in-plane magnetization distribution ($m_x$ and $m_y$) is depicted by white arrows. The white arrows illustrate how the magnetization of the closure domains and Bloch-line arrange at different depths of the slab. **(n-r)** Detail the field evolution from an ordered skyrmions to disordered skyrmions.

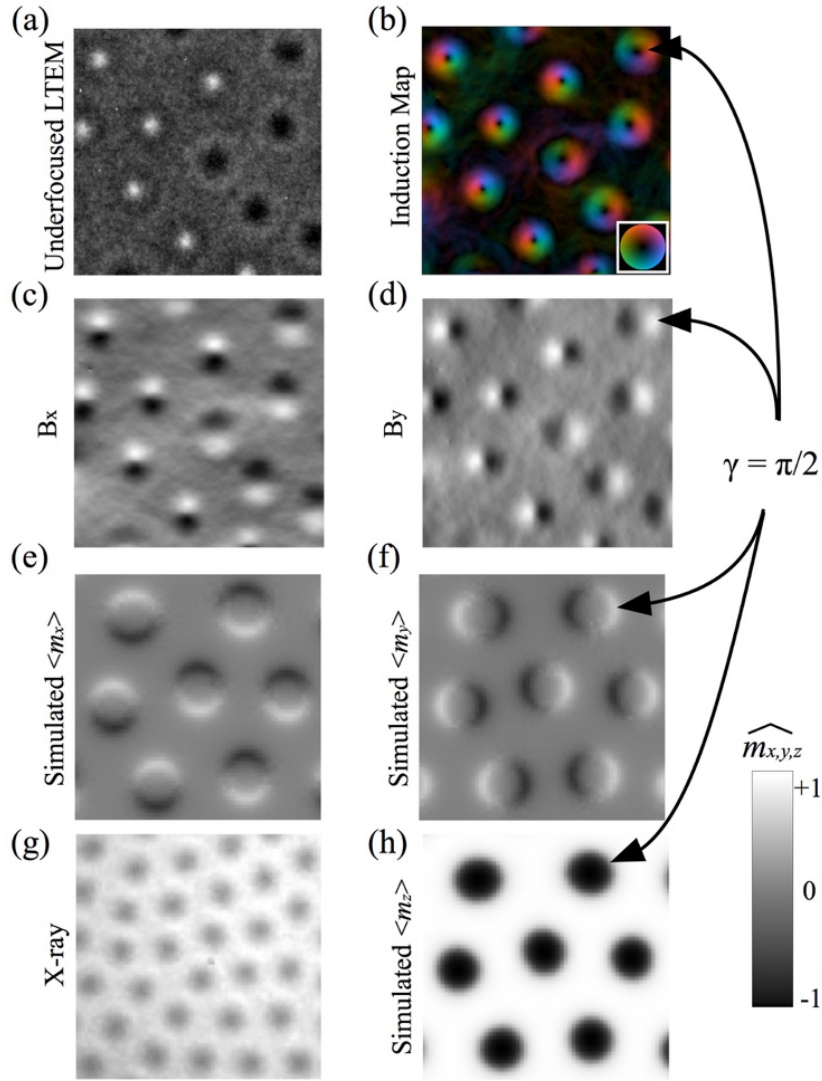

**Figure 6. Comparison between numerical and experimental observation of skyrmions. (a, b)** Under-focused Lorentz-TEM image and its corresponding magnetic induction map illustrate the Bloch-line arrangement of the skyrmions. **(c, d)** The individual in-plane components of the magnetic induction $B_x$ and $B_y$ are obtained from **(b)**. **(e, f)** Detail a numerical computed average magnetization across the slab for $<m_x>$ and $<m_y>$. **(g)** A transmission X-ray microscopy of the skyrmion phase solely shows the presence of cylindrical domains which resembles the average magnetization across the slab for $<m_z>$ in **(h)**. A skyrmion with $S = 1$, $\gamma = \pi/2$ is referenced to directly compare experimental and numerical results in **(b, d, f, h)**. The scale bar references the micromagnetic domain states.

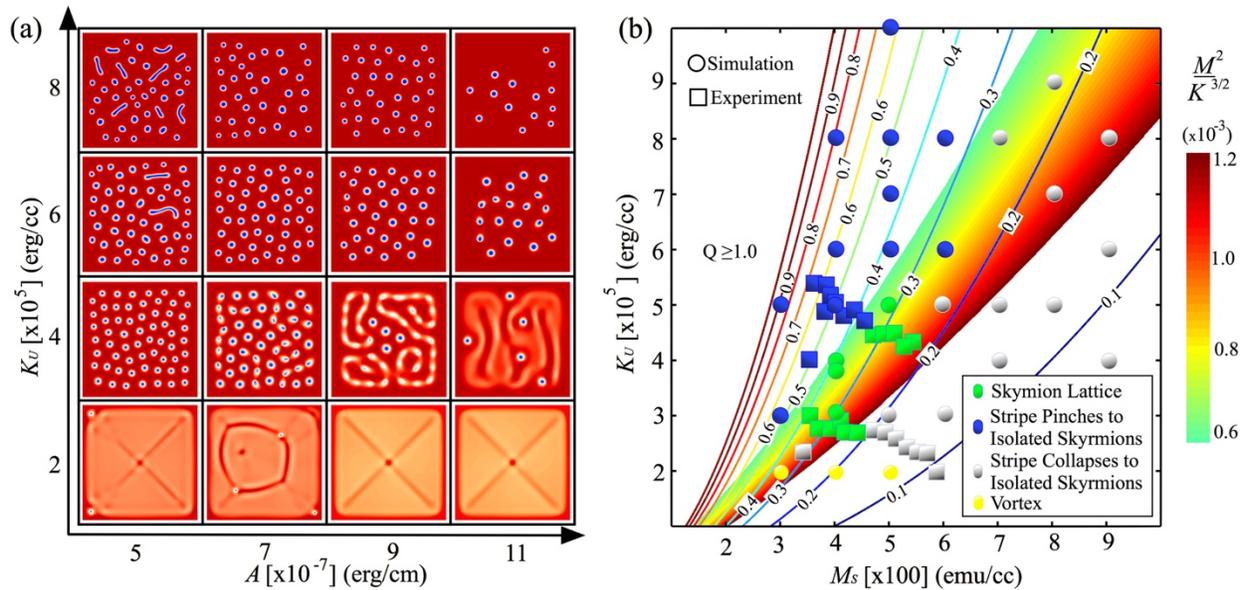

**Figure 7. Skyrmion phase stability in magnetic property phase maps. (a)** The top side view of equilibrium states that form at $H_z$ = 2000 Oe are investigated under different ratios of $K_U$ and $A$ for a fixed $M_S$ = 400 emu/cm$^3$. These equilibrium states detail the magnetization along the z-axis ($m_z$) at the top surface of the slab (z=40nm). The magnetization ($m_z$) is represented by regions in red (+$m_z$) and blue (-$m_z$); whereas the in-plane magnetization ($m_x$, $m_y$) is represented by white regions surrounding the blue features. **(b)** The formation of a skyrmion lattice is explored in terms of varying $M_S$ and $K_U$ for a fixed $A = 5\times10^{-7}$ erg/cm using a combination of experimental results (□) and simulations (○). Contour lines detail the corresponding $Q$-factor in this magnetic phase region, which denotes the balance between anisotropy energy $K_U$ and magneto-static energy $2\pi M_S^2$. For details pertaining to the simulation parameters, see Methods.

# Supplementary Information

# Tailoring magnetic energies to form dipole skyrmions and skyrmion lattices

S. A. Montoya[1,2], S. Couture[1,2], J. J. Chess[4], J. C. T. Lee[4,5], N. Kent[7], D. Henze[7], S. K. Sinha[3], M.-Y. Im[5,6], S. D. Kevan[4,5], P. Fischer[5,7], B. J. McMorran[4], V. Lomakin[1,2], S. Roy[5], and E. E. Fullerton[1,2,**]

[1]Center for Memory and Recording Research, University of California, San Diego, La Jolla, CA 92093, USA
[2]Department of Electrical and Computer Engineering, University of California, San Diego, La Jolla, CA 92093, USA
[3]Department of Physics, University of California, San Diego, La Jolla, CA 92093, USA
[4]Department of Physics, University of Oregon, Eugene OR 97401, USA
[5]Center for X-ray Optics, Lawrence Berkeley National Laboratory, Berkeley CA 94720, USA
[6]Department of Emerging Materials Science, DGIST, Daegu, Korea
[7]Physics Department, University of California, Santa Cruz, CA 94056, USA

### A. Resonant soft X-ray scattering.

Figure 1 shows diffraction images exemplifying the different magnetic phases that are observable in [Fe (0.34nm) / Gd (0.4nm)]x80 along the Gd $M_5$ (1198eV) absorption edge and Fe $L_3$ (708eV) absorption edge. The disordered stripe domains are characterized by a ring-like scatter pattern (Fig. 1a) which agrees with our observation of disordered domains in Lorentz TEM and full-field transmission X-ray microscopy. Figure 1b shows the typical diffraction image obtained from the coexistence of stripes and skyrmions which depicts evidence of 6 Bragg peaks where two peaks exhibit higher intensity than the other four peaks. The skyrmion phase is depicted by a six-fold scatter pattern where the Bragg peaks possess near equal intensity indicating the presence of a hexagonal lattice of cylindrical-like magnetic features (Fig. 1c). Last, the uniform magnetization domain is depicted by a flat diffraction image, where there is no evidence of magnetic contrast (Fig. 1d).

### B. Magnetic hysteresis and ferromagnetic resonance measurements.

Inspection of the magnetic hysteresis of [Fe (0.34 nm)/Gd (0.4 nm)]x80and [Fe (0.36 nm)/Gd (0.4 nm)]x80 reveal the loops are characteristic of a film exhibiting perpendicular magnetized stripe-like magnetic domain structure at room temperature [25,27,28,37] (Fig. 2a, b). This agrees with our experimental observation of disordered stripe domains at zero-field. A textbook example of a stripe-domain hysteresis loop occurs in [Fe (0.36 nm)/Gd (0.4 nm)]x80 at room temperature (Fig. 2a) which depicts a small hysteresis around the saturation field. This hysteresis is associated with the mechanism in which stripe domains form and collapse from cylindrical domains.

---

[*] Corresponding author: efullerton@ucsd.edu

The field-dependent behavior of stripe domains is as follows: At zero field, the competition between long range dipolar and short range exchange interactions results in a meander of domains of equal domain size. When a perpendicular magnetic field is applied, the stripe with magnetization opposite will shrink and the other will grow. Eventually the extremities of the stripe domains shrinking will collapse into cylindrical domains, termed magnetic bubbles. At this point, these textures are randomly sparse throughout the film and they tend not to interact with each other due to dipolar field interactions. When the magnetic field is increased further, the bubble domain's diameter decreases until they collapse at a critical field, known as the collapse field, which occurs close to magnetic saturation. In the case the magnetic field is reduced from magnetic saturation, magnetic bubbles will form at pinning (defect) sites and will continue to grow as the field is reduced until reaching a nucleation field, another critical field, which is typically depicted by a sharp kink in the magnetization (Fig 2a). At this field the bubbles are submitted to an elliptical instability [28] which results in the bubble domains branching out to elongated stripe-like domains. By further decreasing the field, the thin film is filled with maze-like domains by means domain wall motion [25,27,28]. In overall, the domain morphology consists of alternating perpendicular magnetic domains with magnetization pointing along/opposite the direction of the magnetic field. In overall, the stripe-to-bubble and bubble-to-stripe mechanism is described by the hysteresis around the nucleation field and this mechanism.

As the temperature is reduced, in Figure 2a, the hysteresis around the nucleation field becomes smaller and softer and the nucleation field is shifted to higher magnetic fields until no identifiable nucleation field exists in [Fe (0.36 nm)/Gd (0.4 nm)]x80. This suggests that the dipolar interaction strength is increasing as the temperature is reduced. Here, our macroscopic magnetic profile agrees with our experimental observation of disordered stripe domains from room temperature down to 200K in [Fe (0.36 nm)/Gd (0.4 nm)]x80. When the temperature is reduced further, the magnetic hysteresis loop appears to suggest the film has undergone a spin-reorientation transition (SRT), yet a closer inspection reveals the films potentially exhibit perpendicular magnetic domains. To verify the orientation of the magnetization, we performed in-plane field magnetic hysteresis measurement and found that the magnetization appears to begin undergoing a temperature driven spin-reorientation transition of the magnetization as we decrease the temperature (Fig 3a); however, the magnetization never fully goes in-plane. The in-plane magnetic hysteresis is characteristic of thin films exhibiting stripe domains - they detail that the magnetization lies on both the perpendicular and along the plane of the thin film. If the [Fe (0.36 nm)/Gd (0.4 nm)]x80 had undergone an SRT of the magnetization, then we should expect the in-plane loops to become square-like and have a negative switching field $H_k$ in this measurement geometry. As the temperature is reduced, the switching field $H_k$ decreases rapidly (Fig. 3b), but does not become negative. Even at 20K, the [Fe (0.36 nm)/Gd (0.4 nm)]x80 shows evidence of perpendicular magnetic domains. This is also observable in the temperature dependence of the intrinsic anisotropy (Fig. 4, main text). Similar observations can be made for [Fe (0.34 nm)/Gd (0.4 nm)]x80. At temperatures where [Fe (0.36 nm)/Gd (0.4 nm)]x80 and [Fe (0.34 nm)/Gd (0.4 nm)]x80 exhibit a skyrmion phase their perpendicular magnetic hysteresis loops are comparable with the exception of different switching fields.

In the case of Fe/Gd bilayer repetitions are varied, we find there is a transition from in-plane magnetization state to a perpendicular magnetization state as the number of bilayers are increased (Fig. 2c). The film studied is [Fe (0.34nm) /Gd (0.41nm)]$_{xN}$ where the repetitions N include 40, 80 and 120. At 40 repetitions, there hysteresis clearly shows the magnetization prefers lying along the plane of the film. Above 40 repetitions, the Fe/Gd appears to build sufficient magnetostatic energy to force the system to break into perpendicular magnetic domains. We note the magnetic hysteresis is similar to those observed in [Fe (0.36 nm)/Gd (0.4 nm)]x80 and [Fe (0.34 nm)/Gd (0.4 nm)]x80 when a skyrmion phase is present. Then, by further increasing bilayer repetitions, we find the [Fe (0.34nm) /Gd (0.41nm)]x120 shows a magnetic hysteresis loop that also suggests a domain morphology with perpendicular magnetized stripe-like magnetic domains.

Cape and Lehman [27] have calculated the magnetization curves for both the bubble-lattice array and the parallel-stripe array structures and show that both ground states can exhibit near identical magnetic hysteresis profiles. Among the differences, is the slope and shift of the nucleation and collapse critical fields. For instance, the nucleation field of the stripe phase is given by a sharp kink of the magnetization, as detailed previously; whereas, the bubble-lattice phase will show a softer slope change in magnetization around the same nucleation field like that observed in [Fe (0.34nm) /Gd (0.41nm)]x120 (Fig 2c). All in all, the magnetization curves provide a good macroscopic depiction of the domain morphologies present in the thin films detailed.

To determine required anisotropy at which skyrmions form, we performed ferromagnetic resonance measurements in fixed frequency mode while a perpendicular magnetic field is scanned. By looking at the resonance resulting from the homogenous magnetization precession, known as the Kittel resonance, we can determine the effective magnetization of the specimen. All Fe/Gd films exhibit a positive effective magnetization $4\pi M_{eff}$ which indicates they possess weak perpendicular magnetic anisotropy (Fig. 1d, e). Both [Fe (0.34 nm)/Gd (0.4 nm)]x80 and [Fe (0.36 nm)/Gd (0.4 nm)]x80 show a linear increase of the effective magnetization $4\pi M_{eff}$ as the temperature is reduced (Fig. 2d, e) which suggests the film becomes more in-plane as the temperature is reduced. For the [Fe (0.34nm) /Gd (0.41nm)] films, we only detail the Kittel resonance at room temperature for all three different bilayer structures. We find that $4\pi M_{eff}$ deviates between 40 repetitions and higher number of bilayer repetitions (Fig. 2f). This suggests a microstructural change occurred during the deposition of the 40 repetition stack which resulted in a slightly different film; whereas, both 80 and 120 repetitions have very close effective magnetization values.

## C. Small angle x-ray reflectivity

Small angle x-ray reflectivity (XRR) measurements were performed on the Fe/Gd films to determine accurately the thickness of the Fe/Gd multilayers and to resolve information regarding the bilayers (Fig. 3). From the XRR data, we find two oscillating waves superimposed that correspond to the Fe/Gd multilayers (fast oscillations) and the Ta capping layer (slow oscillations). By fitting the XRR data to the modified Bragg law, we find both films are thinner than the expected deposited thickness – [Fe (0.36 nm)/Gd (0.4 nm)]x80 is 51.7 nm and [Fe (0.34 nm)/Gd (0.4 nm)]x80 is 53.6 nm – which suggests the deposited ultra-thin Fe and Gd layers intermix. Both Fe/Gd films show no evidence of super-lattice peaks at any 2θ angles. This suggest the films do not have well defined bilayers throughout the film thickness. Given the individual layers are each a few monolayers thick, strong intermixing between the layers can be expected. As a result, the thin film likely consist of continuously varying alloy-like layers.

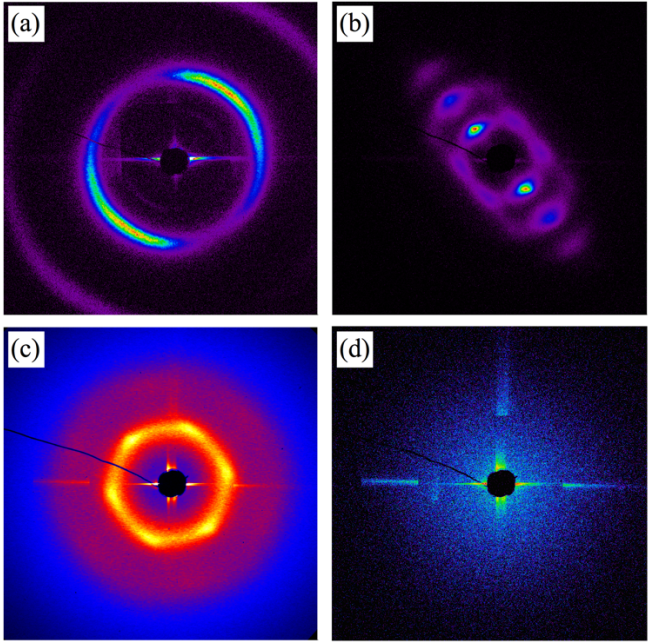

**Figure 1. Reciprocal space imaging of the field-dependent magnetic domain morphology of [Fe (0.34 nm)/Gd (0.4 nm)]x80.** The scatter images obtained at room temperature detail the four magnetic phases observable using this technique: **(a)** disordered stripe domains, **(b)** coexisting stripes and skyrmions, **(c)** skyrmion lattice and **(d)** uniform magnetization. **(a)** The diffraction image is obtained along the Fe $L_3$ (708eV) absorption edge at zero-field at 85K. **(b, c)** These diffraction images are both obtained at room temperature along the Gd $M_5$ (1180eV) absorption edge at **(b)** $H_z$ = 1500 Oe and **(c)** $H_z$ = 1900 Oe. **(d)** The saturated state is obtained along the Fe $L_3$ (708eV) absorption edge at $H_z$ = 5000 Oe at room temperature.

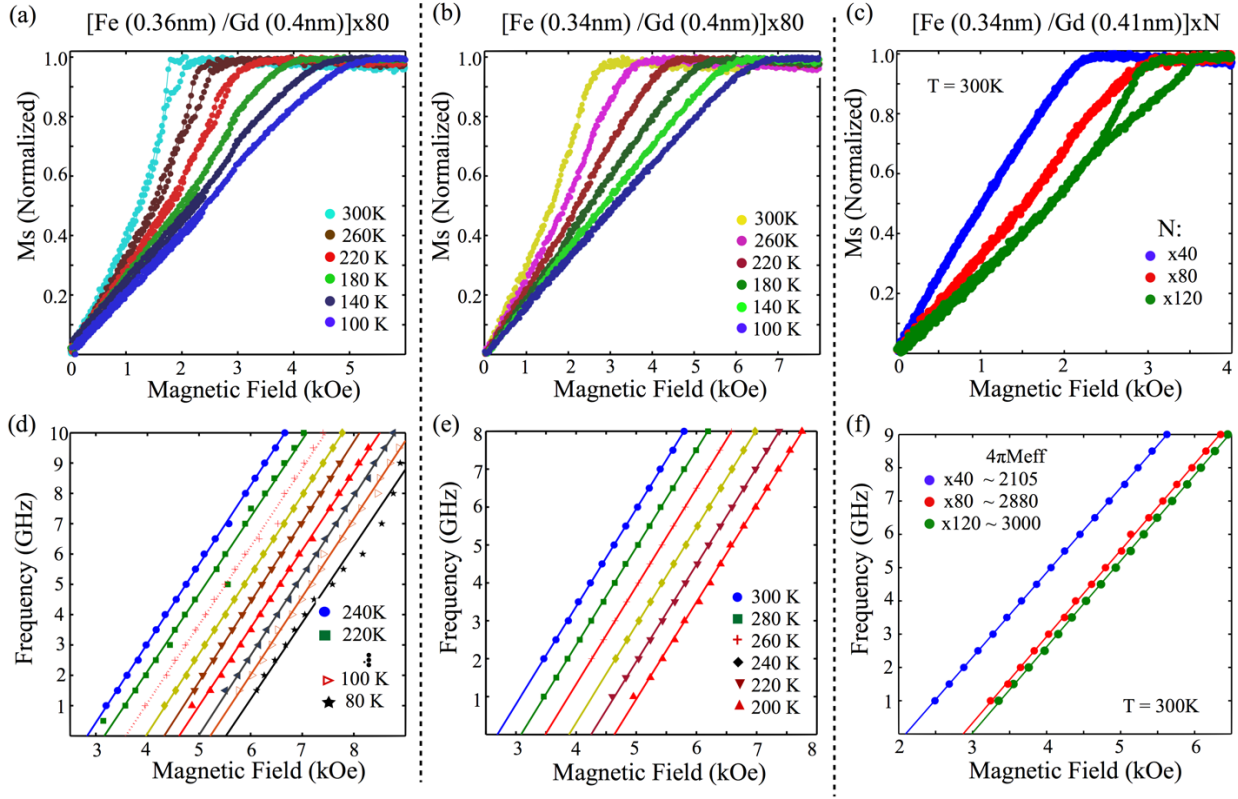

**Figure 2. Magnetic properties of Fe/Gd films. (a, b, c)** The magnetic hysteresis loops and **(d, e, f)** the Kittel resonance are detailed for Fe/Gd films presented in this work. **(a, b, c)** Show the temperature dependent magnetic hysteresis loops for [Fe (0.36nm)/Gd (0.4nm)]x80, [Fe (0.34nm) /Gd (0.4nm)]x80 and [Fe (0.34nm) /Gd (0.41nm)]xN where N is 40, 80 and 120 repetitions. The films are saturated with a magnetic field of ±10 kOe, but for ease of viewing a smaller field range is illustrated. **(d, e, f)** Depict the field-dependent resonance resulting from the precession of the homogenous magnetization for the Fe/Gd films in **(a, b, c)**.

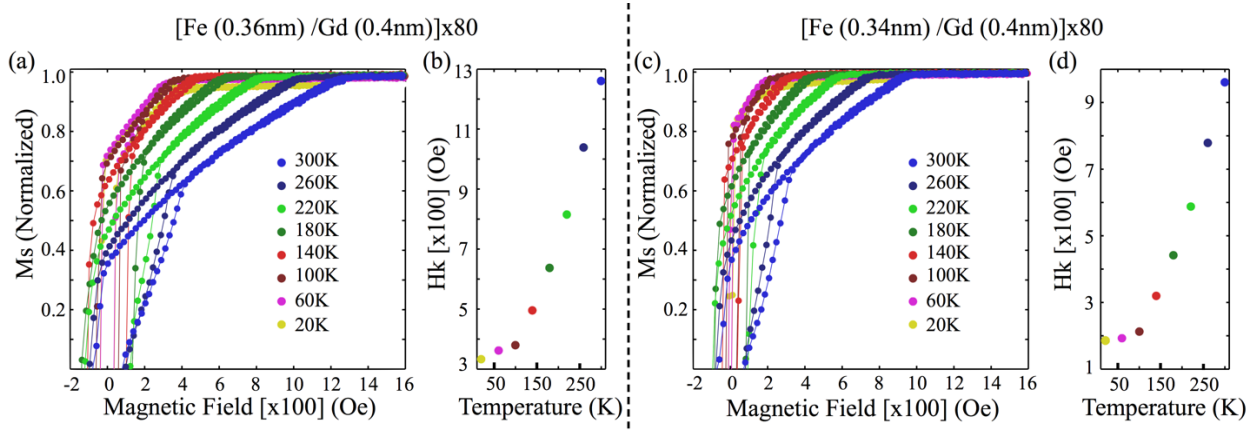

**Figure 3. In-plane magnetic properties. (a, c)** The temperature dependent normalized in-plane magnetic hysteresis loops are detailed for [Fe (0.36nm) /Gd (0.4nm)]x80 and [Fe (0.34nm) /Gd (0.4nm)]x80. From **(a, c)** the temperature dependent switching field, $H_k$, is extracted and is illustrated in **(b, d)**. As the temperature is reduced, the switching field decreases swiftly but does not become negative as expected for a perpendicular magnetized films that is characterized by a square hysteresis loop. The films are saturated with a magnetic field of ±10 kOe, but for ease of viewing a smaller field range is illustrated.

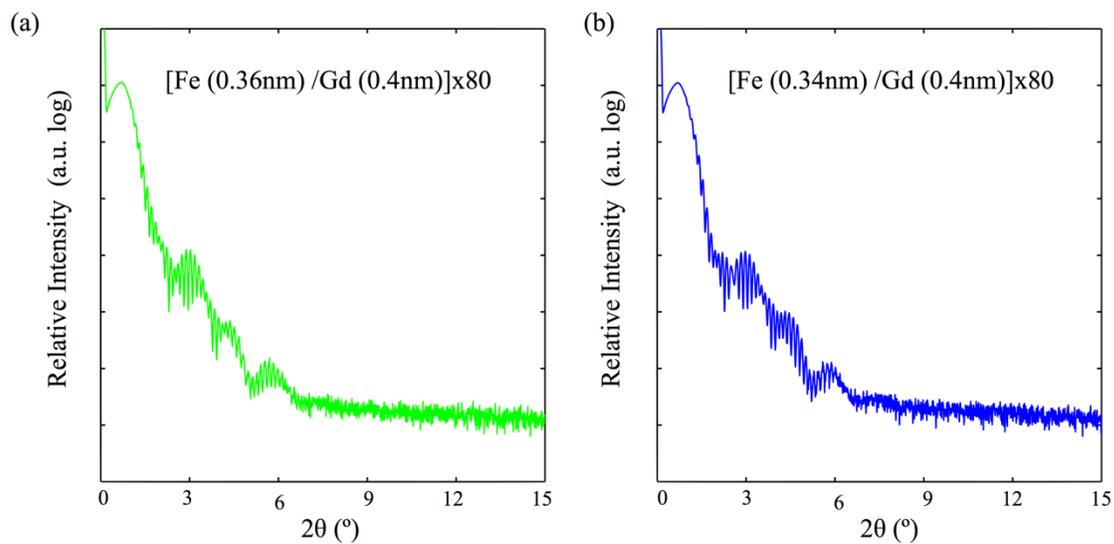

**Figure 4. X-ray Reflectivity.** The small angle reflectivity for **(a)** [Fe (0.36 nm)/Gd (0.4 nm)]x80 and **(b)** [Fe (0.34 nm)/Gd (0.4 nm)]x80 are detailed from 2θ angles 0º to 15º.